\documentclass[twocolumn]{aastex61}
\usepackage{natbib}
\usepackage{amsmath}

\newcommand{\logg}{\ensuremath{\log g}}

\usepackage{graphicx}
\usepackage{hyperref}

\newcommand\aastex{AAS\TeX}

\received{\today}
\shorttitle{\aastex\ Back to the Lithium Plateau}
\shortauthors{Aguado et al.}
\begin{document}
   \title{Back to the Lithium Plateau with the [Fe/H] $<$ -6 star J0023+0307\footnote{
Based on observations made with Very Large Telescope (VLT) at Paranal Observatory, Chile}
 }

\correspondingauthor{David~S. Aguado}
%\email{aguado@iac.es, jonay@iac.es, callende@iac.es, rrl@iac.es}
\email{aguado@iac.es}

%\author{IAC EMP GROUP}
\author[0000-0001-5200-3973]{David~S. Aguado}
\affil{Instituto de Astrof\'{\i}sica de Canarias,
            V\'{\i}a L\'actea, 38205 La Laguna, Tenerife, Spain\\}
\affiliation{Institute of Astronomy, University of Cambridge, Madingley Road, Cambridge CB3 0HA, UK \\}

\author[0000-0002-0264-7356]{Jonay~I. Gonz\'alez~Hern\'andez}
\affil{Instituto de Astrof\'{\i}sica de Canarias,
              V\'{\i}a L\'actea, 38205 La Laguna, Tenerife, Spain\\}
 \affiliation{Universidad de La Laguna, Departamento de Astrof\'{\i}sica, 
             38206 La Laguna, Tenerife, Spain \\} 
\author[0000-0002-0084-572X]{Carlos Allende Prieto}
\affil{Instituto de Astrof\'{\i}sica de Canarias,
              V\'{\i}a L\'actea, 38205 La Laguna, Tenerife, Spain\\}
 \affiliation{Universidad de La Laguna, Departamento de Astrof\'{\i}sica, 
             38206 La Laguna, Tenerife, Spain \\} 
\author[0000-0003-3767-7085]{Rafael Rebolo}
\affil{Instituto de Astrof\'{\i}sica de Canarias,
              V\'{\i}a L\'actea, 38205 La Laguna, Tenerife, Spain\\}
 \affiliation{Universidad de La Laguna, Departamento de Astrof\'{\i}sica, 
             38206 La Laguna, Tenerife, Spain \\}             

\affiliation{Consejo Superior de Investigaciones Cient\'{\i}ficas, 28006 Madrid, Spain\\}

\begin{abstract}
We present an analysis of the UVES high-resolution spectroscopic observations at the 8.2m VLT of J0023+0307, a main-sequence extremely iron-poor dwarf star. We are unable to detect iron lines in the spectrum but derive [Fe/H]$<-6.1$ from the Ca II resonance lines assuming [Ca/Fe]$\geqslant0.40$.
The chemical abundance pattern of J0023+0307, with very low [Fe/Mg] and [Ca/Mg] abundance ratios, but relatively high absolute Mg and Si abundances, suggests J0023+0307 is a second generation star formed from a molecular cloud polluted by only one supernova in which the fall-back mechanism played a role. We measure a carbon abundance of A(C)$=6.2$ that places J0023+0307 on the ``low'' band in the A(C)$-$[Fe/H] diagram, suggesting no contamination from a binary companion. 
This star is also unique having a lithium abundance (A(Li)$=2.02\pm0.08$) close to the level of the Lithium Plateau, in contrast with lower Li determinations or upper limits in all other extremely iron-poor stars. The  upper envelope of the lithium abundances in unevolved stars  spanning more than three orders of magnitude in metallicity ($-6<$[Fe/H]$<-2.5$) defines a nearly constant value. We argue that it is unlikely that such uniformity is the result of depletion processes in stars from a significantly higher initial Li abundance, but suggests instead a lower primordial production, pointing to new physics such as decaying massive particles, varying fundamental constants, or nuclear resonances, that could have affected the primordial $^7$Li production.
\end{abstract}


\keywords{stars: Population II --- stars: individual (J0023+0307)---   Galaxy: formation --- Galaxy: halo ---  Cosmology: observations  --- primordial nucleosynthesis}

\section{Introduction} \label{intro}
The study of most ancient stars in the Milky Way allows us to infer early properties of the Galaxy, its chemical composition, and assembly history. The number of ultra metal-poor stars (i.e., [Fe/H]$<-4.0$) has increased strikingly in the past few years. A large observational effort has enlarged the sample of such rare objects, thanks to new search techniques 
\citep[see e.g.,][]{chris01,fre05,caff11,kel14,boni15,han15II,sta18}. 
The main sources of metal-poor candidates are large spectroscopic surveys such as Hamburg/ESO \citep{chris01}, SDSS \citep{yor00} or LAMOST \citep{Deng12}, and photometric ones like Skymapper \citep{kell07} or Pristine \citep{sta17}. 

Metal-poor stars are invaluable messengers that carry information from early epochs and an important key to understand the primordial production of lithium
and the processes responsible for the possible ``meltdown''~\citep[see e.g.,][]{aoki09,sbo10} of the Lithium plateau~\citep{spi82,reb88}. 
All the stars with metallicities below [Fe/H]$<-3.0$ and lithium abundances lower than A(Li)$=2.2\pm0.1$ are considered as likely affected by destruction in the stars. Several explanations have been discussed~\citep[see e.g.,][]{rich05,piau06,jon09b,mel10,sbo10,mol12,boni15,mat17II}, including lithium depletion in stellar atmospheres due atomic diffusion, high rotational velocities during the earlier star formation, astration by population III stars, destruction in the pre-main sequence phase, the presence of enhanced fragmentation in the formation of ultra metal-poor stars.

New or poorly measured nuclear reaction resonances could affect the lithium production predicted in Big Bang Nucleosynthesis (BBN) \citep[e.g.,][]{cyb12}, although recent experiments make this nuclear ``fix" to the BBN unlikely~\citep{oma11}. Processes injecting neutrons at the relevant temperatures of the primordial plasma can also alter the primordial $^7$Li abundance, for instance, decaying massive particles (well-motivated candidates such as weakly interacting massive particles (WIMPs) are provided by supersymmetry) are a natural source for a neutron excess~\citep{pes08} which injected at the appropriate level may reduce the primordial production of $^7$Li. In addition, time-varying fundamental constants may lead to a significant lower value for primordial $^7$Li, which is particularly sensitive to changes in the deuteron binding energy \citep[see e.g.,][]{fie11}. Lithium observations in stars at the lowest observed metallicities are specially important to bring insight on the processes of potential lithium depletion in stars and, ultimately, to establish if any non-standard physics may have played a role during or after BBN.

J0023$+$0307 was discovered and firstly observed by \citet{agu18II}. \citet{pat18} observed it with X-shooter on the 8.2\,m~VLT, providing for first time individual abundances of carbon, magnesium and silicon. Most recently, using MIKE on the Magellan telescope, \citet{fre18} added additional elemental abundances and claimed a lithium detection. Here, we measure with high precision many elemental abundances that confirm most of the results from the literature and provide a reliable lithium abundance determination of J0023$+$0307. These observations allow us to explore the lower-metallicity end of the A(Li)$-$[Fe/H] diagram. 

\section{Observations} \label{obs}
J0023$+$0307 (g=17.90$\pm 0.01$, R.A.$=00^{h} 23^{m} 14^{s}.00$, DEC.=$+03^{0} 07^{'} 58^{''}.07$ (J2000)) was observed under the ESO program 0101.D-0149(A) with the UVES spectrograph on the 8.2\,m Kueyen Very Large Telescope at Cerro Paranal Observatory, in the Atacama desert, Chile. The observations were taken in twelve observing blocks (OB's), one hour each on 2018 July 22-23, Aug 8-12, and Oct 7-8. The total on-target integration time was 10 hours. A $1\farcs4$ slit was used with 2x2 binning in dark sky conditions. Our settings used dichroic $\#Dic1$ and provided a spectral coverage between 390 and 580~{nm}. The data were reduced by the ESO staff, and retrieved through the Phase 3 query form. We corrected each spectrum for the barycentric velocity, coadded and combined them.
The signal-to-noise per 0.018{\AA}/pixel in the combined spectrum was $\sim$30, 55 and 80 at 380, 510 and 670~{nm} respectively.
The nominal resolving power for this set up is R$\sim31,220$ for the blue part of the spectrum ($330-452$~{nm}) and R$\sim31,950$ for the red ($480-680$~{nm}). The seeing was $0\farcs91$ during the first OB and between $0\farcs40$ and $0\farcs65$ during the rest of the run, thus the actual $R$ value is somewhat higher. 
We performed a fit using an automated fitting tool based on IDL MPFIT routine (with continuum location, global shift, abundance, and global FHWM as free parameters) of all individual, isolated, detected spectral lines (see Table~\ref{lines}) with a high-resolution model and obtained an average global (including instrumental and macroturbulent broadening and assuming no rotation) Gaussian broadening of $7.4\pm0.6$\,km s$^{-1}$~(equivalent to $R\sim$40,500).
From a cross-correlation of the J0023$+$0307 spectrum with a extremely metal-poor (EMP) F-type star template we derive a radial velocity of $v_{rad}= -195.5\pm 1.0$\,km s$^{-1}$, in perfect agreement with the value reported by \citet{fre18}, $v_{rad}= -194.6\pm 1.2$\,km s$^{-1}$. No signal of $v_{rad}$ variation has been observed in the observations spanning four months, suggesting the star may be single. 
The Gaia DR2 \citep{gaia2018} provides a parallax of $\varpi=0.27\pm 0.14$\,mas. 
A detailed calculation of the Galactocentric orbit has been recently carried out by \citet{ses18}, who find that the orbit of J0023$+$0307 is confined to the Galactic plane within $|Z|<2.3$kpc.

\section{Stellar parameters}\label{anali}
\begin{figure*}[!ht]
\begin{center}
{\includegraphics[clip=true,width=85 mm, angle=180]{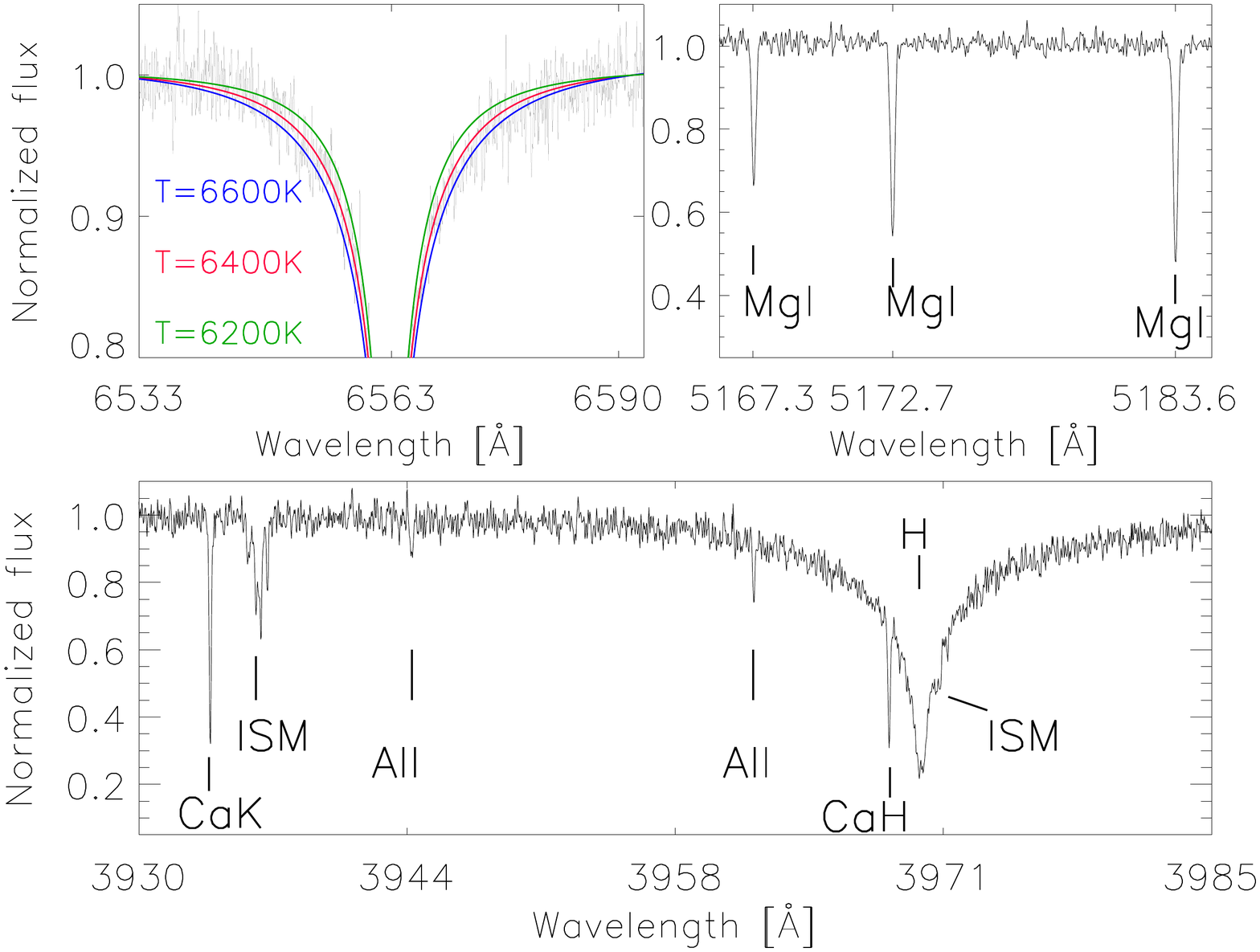}}
{\includegraphics[clip=true,width=85 mm, angle=180]{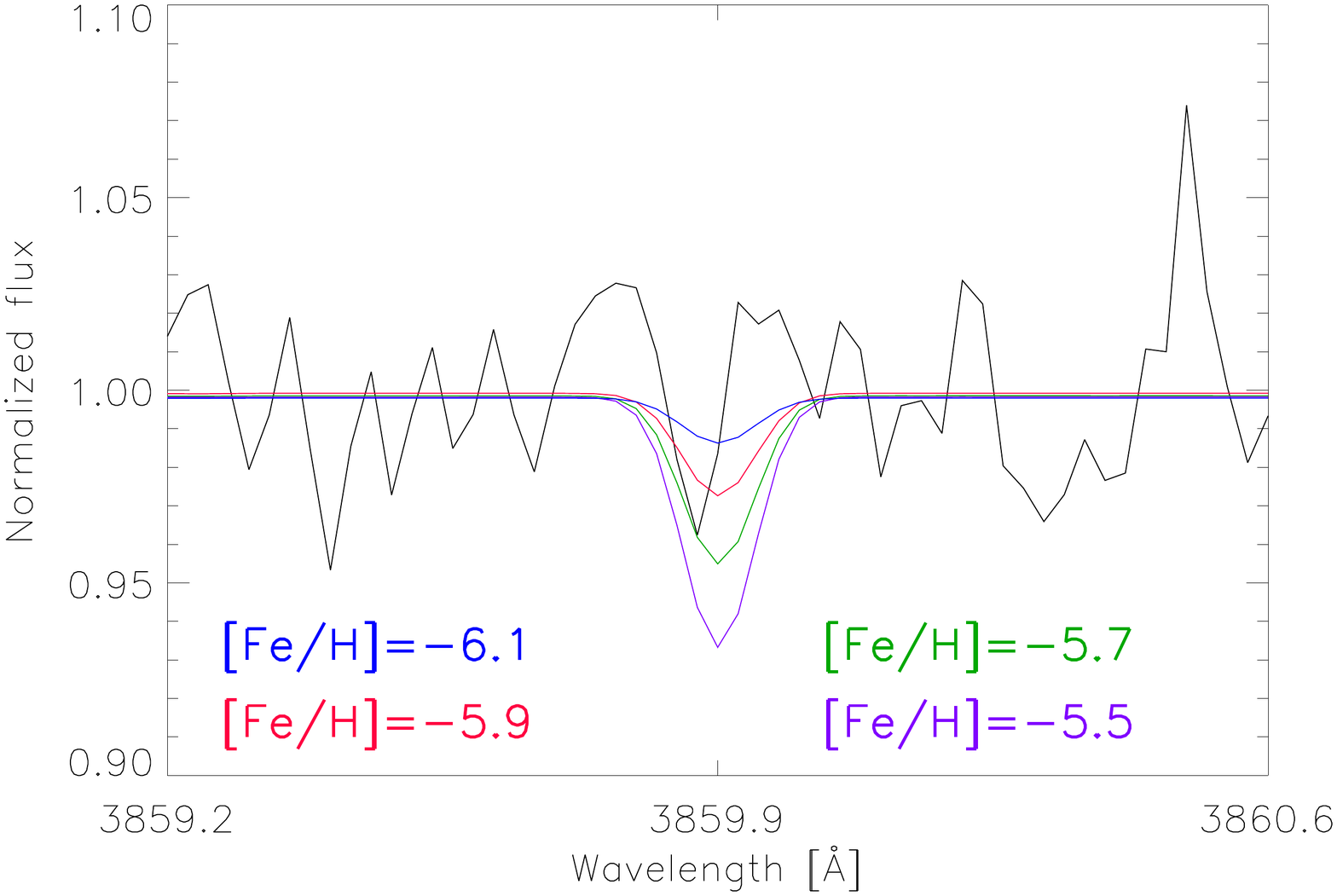}}
{\includegraphics[clip=true,width=85 mm, angle=180]{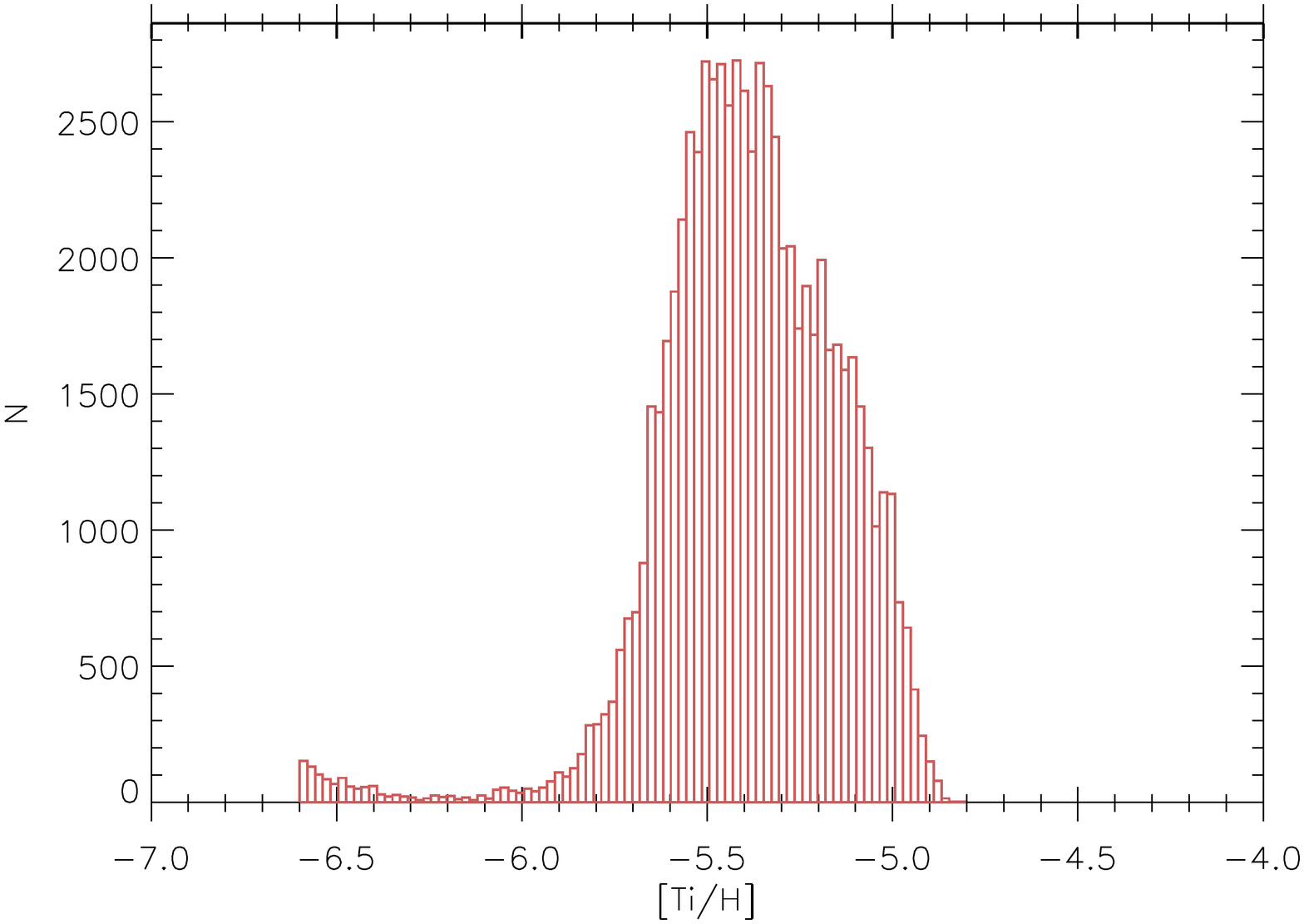}}
{\includegraphics[clip=true,width=85 mm, angle=180]{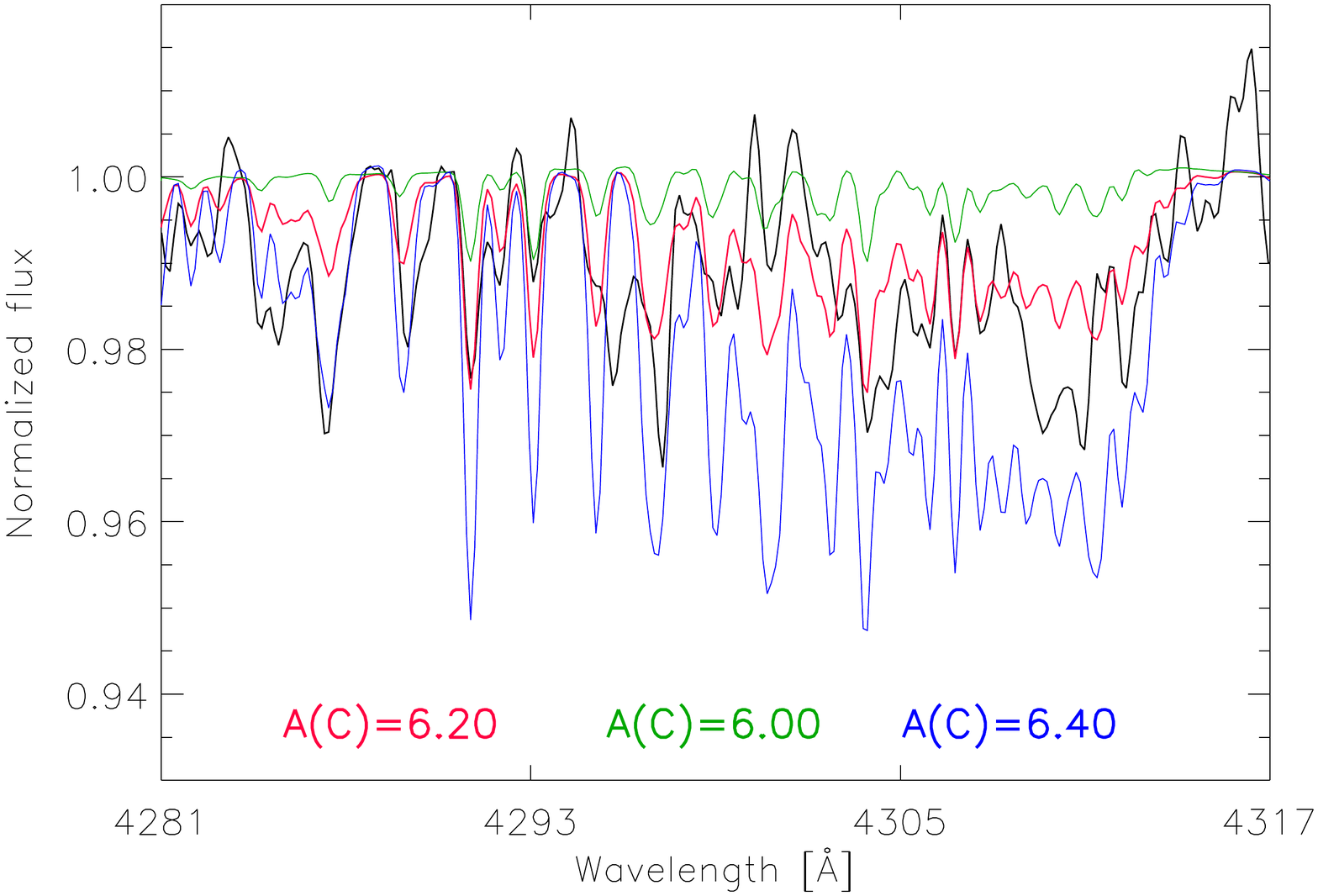}}
\end{center}
\caption{\textit{Upper-left panel}: The first sub-panel shows the UVES J0023$+$0307 spectrum (black line) in the H$_{\alpha}$ region around 6563\,\AA\ together with the best fit (red line) and the upper and lower limit values (blue and green lines). For fitting purposes the H$_{\alpha}$ core has been excluded from the $\chi^{2}$ calculation. The other two sub-panels show a detail of the J0023$+$0307 UVES spectrum in the vicinity of the magnesium triplet and calcium H\&K.
\textit{Upper-right panel}: A detail of the J0023$+$0307 UVES spectrum in the vicinity of the strongest iron line at 3859.9\,\AA. A synthetic spectra of four different iron ratios ([Fe/H]=-6.1, -5.9, -5.7, -5.5) are plotted.
%Middle-left panel: A detail of two titanium features with (black line) and the best fit derived with FERRE fitting at the same time 15 titanium lines in the blue. For reference two synthetic spectra are shown, [Ti/H]=-5.8, -4.9, in green and blue respectively. 
\textit{Lower-left panel}: The number of Markov Chain Monte Carlo experiments vs [Ti/H] (15 lines) computed with FERRE. 
\textit{Lower-right panel}: 
%The G-band area of the UVES J0023+0307 spectrum (black line), smoothed to 10,000 resolution, compared with the best fit derived by FERRE (red line) after apply a running mean filter normalization with a filter of 200-pixel window together with a upper and lower values (blue and green lines).
The G-band area of the UVES J0023+0307 spectrum (black line), smoothed to 10,000 resolution, compared with ASSET synthetic spectra for the best fit carbon abundance (red line) together with a upper and lower values (blue and green lines).
}
\label{carbon}
\end{figure*}

\citet{agu18II} reported effective temperatures from the analysis of BOSS ($T_{\rm eff}=6295 \pm 36$\,K), OSIRIS ($T_{\rm eff}=6140 \pm 132$\,K), and ISIS ($T_{\rm eff}=6188\pm 84$\,K) low/mid-resolution spectra. After that, \citet{pat18} derived a value of $T_{\rm eff}=6160 \pm 100$\,K from $(g-z)$ color calibrations. By fitting the wings of the $H_{\alpha}$ Balmer line we find a higher temperature ($T_{\rm eff}=6400 \pm 150$\,K) (See Fig. \ref{carbon}). In addition, we used the color calibrations based on the infrared flux method presented in \citet{jon09} with $2MASS$ and $V$ Johnson magnitudes transformed from UKIDSS $J,H,K$ and SDSS $g,r$ filters. This approach also led to a higher temperature ($T_{\rm eff}=6482 \pm 224$\,K). On the other hand, using color-temperature-metallicity calibrations by \citet{cas10} we arrive at three very different temperatures from different colors $V-J$, $V-H$ and $V-K$, giving an average $T_{\rm eff}=6441\pm 158$\,K.

\begin{table}
\begin{center}
%\centering
\caption{\label{tempe}All the T$_{\rm eff}$ derived values considered in this work and explained in Section \ref{anali}}
\begin{tabular}{llrrrl}
\hline
  Source & Ref. & T$_{\rm eff}$  & $\delta{T}$ & \\
\hline
\hline
BOSS spectrum   & (1) & 6295 & 36 & \\
OSIRIS spectrum & (1) & 6140 & 132&  \\
ISIS spectrum   & (1) & 6188 & 84 & \\
(g-z)           & (2) & 6160 & 100 &  \\
H$_{\alpha}$    & (3) & 6400 & 150 & \\
H$_{\beta}$     & (3) & 6165 & 62 &\\
IRFM            & (4) & 6482 & 224 & \\
(V-J)           & (5) & 6481 & 156 & \\
(V-H)           & (5) & 6335 & 186 & \\
(V-K$_{s}$)     & (5) & 6615 & 212 & \\
\hline
Mean  V-X       & (5) & 6474 & 145 & \\
\hline
(V-I)           & (5) & 5992 & 157 & \\
(V-I)           & (6) & 5997 & 130 &\\
\hline
\hline
%\tablefoot{References: (1) \citet{agu18II}, (2) \citet{pat18}, (3) This work, (4) \citet{jon09}, (5) \citet{cas10},  (6) \citet{fre18}}
\end{tabular}
\end{center}
\end{table}

Finally, \citet{fre18} use a lower value ($T_{\rm eff}=5997 \pm 130$\,K) from the $V-I$ color. The dispersion found along the several methods performed by different authors reflects how difficult is to derive a robust effective temperature in EMP stars~\citep[see e.g.,][]{sbo10}. We decided to adopt the $T_{\rm eff}$ value from \citet{pat18}, $T_{\rm eff}=6160 \pm 100$\,K, which is essentially the same as derived by \citet{agu18II} from a high S/N ISIS spectrum and also very close to the weighted mean temperature of the values given in Table~\ref{tempe}.
%In Table~\ref{tempe} all the different effective temperature determinations are summarized.

The surface gravity derived by \citet{agu18II}, $\logg=4.9 \pm0.5$, confirms that J0023+0307 is a dwarf star. \citet{pat18} assumed $\logg=4.5 \pm0.5$ arguing that the star is not bright enough to be considered a subgiant based on the %distance derived from the Gaia DR2 parallax by \citet{bay18}. In addition, a
distance derived from the Gaia DR2 parallax. In addition, \citet{ses18} carried out a kinematical study and concluded that the subgiant solution implies an unbound orbit, favoring the dwarf against the subgiant solution at 88\% confidence level. Since there are no detectable iron lines in the UVES spectrum of J0023+0307, we adopt the value from \citet{ses18}, $\logg=4.6 \pm0.1$, which is consistent with all other estimates.
%The estimated extinction using the Schlegel et al. (1998) maps is $E(B-V)=$0.0325 using \citet{boni00} corrections. 
Following the method explained in \citet{alle06}, based on the comparison of the main stellar parameters with different sets of isochrones from \citet{ber94} we find a galactocentric distance of $\sim 2.6$\,kpc. The obtained value is in agreement with those derived by \citet{ses18}, d=$ 2.7 \pm 1.4 $\,kpc, using Gaia DR2 data. 

Finally, the microturbulence has been fixed at 2\,km s$^{-1}$. The errors introduced by using this value instead of a value more appropiate for a dwarf of 1.5\,km s$^{-1}$ are small ($<0.1$\,dex), as discussed in \citet{agu18I} and \citet{sta18} and included in the our quoted derived abundances uncertainties in this work. In the metallicity regime [Fe/H]$<-5.0$, the analysis is performed with model atmospheres computed with [Fe/H]$=-5.0$.

\section{Analysis}\label{analisis}
A grid of synthetic spectral models has been computed with the code ASSET \citep{koe08} using stellar models from Kurucz ATLAS 9 \citep{mez12} and following the recipe explained in \citet{alle14} and \citet{agu17I}. The grid contains models spanning $-7\leq[\rm Fe/H]\leq-4$; $-1 \leq[\rm C/Fe]\leq 7$, 
$4750\,\rm K\leq T_{eff}\leq7000\,\rm K$ and $1.0\leq \logg \leq5.0$. 
This grid extends the one used by \citet{sta18} to reach lower metallicities. Both grids have been smoothed to R$\sim40,500$ and normalized. We have performed an MCMC analysis based in self-adaptative randomized subspace sampling \citep{vru09} implemented in the FERRE code\footnote{{\tt FERRE} is available 
from http://github.com/callendeprieto/ferre} \citep{alle06}. The MCMC algorithm samples the probability distribution to estimate central values and uncertainties. Further details about the use of FERRE with high-resolution spectroscopy are presented in \citet{sta18}.

\subsection{Iron and titanium}
Similarly to SMSS J0313$-$6708 \citep{kel14} or J1035$+$0641 \citep{boni15} no iron lines have been detected in the UVES spectrum. We are able to provide an upper limit of [Fe/H]$<-5.5$ (see Fig.~\ref{carbon}, upper-right panel) only from the strongest iron line in the spectrum at 385.9~{nm}. The S/N ratio in the blue region of the spectrum is $\sim35$, where no clear iron features are present. However we have analyzed simultaneously 20 regions of 0.08~{nm} (between 372- and 405~{nm}) corresponding to wavelengths of 20 isolated iron lines, and we have sampled the probability distribution by launching 10 Markov chains of 10,000 experiments each. We arrive at [Fe/H]=$-6.92\pm 0.23$ but we cannot consider this result final due to the proximity of the grid limit and the extremely low signal we are trying to measure. \citet{fre18} performed an analysis using only the Fe line at 385.9~{nm} and provided an upper limit of [Fe/H]$<-5.6$. On the other hand, we have found a more robust and reliable result with the same FERRE-MCMC analysis on 15 Ti lines. In Fig. \ref{carbon} (lower-left panel), we plot the probability distribution from the MCMC experiments (right). With this methodology we are able to recover information hidden in the noise, by performing this analysis simultaneously over 15 Ti lines, arriving to a [Ti/H]=$-5.37\pm 0.15$. \citet{fre18} arrived at [Ti/H]$<-4.3$ (1 line) and [Ti/H]$<-4.5$ (13 lines).

\subsection{Carbon}
The moderate S/N in the G-band region forces us to perform a careful continuum determination. Even in Carbon-Enhanced Metal-Poor (CEMP) stars, especially with $T>6000$\,K, it is hard to distinguish CH features from the noise. We have pre-normalized the J0023+0307 UVES spectrum using a low-order polynomial and then normalized both the observed and synthetic spectra using a running mean filter with a 200 pixel window. Since we are only comparing neighbouring pixels, i.e., relative fluxes, our methodology partially solves the problem of the continuum determination. The best FERRE fit leads to A(C)$=6.20\pm0.2$. An additional test was done after smoothing the spectrum to $R\sim10,000$ resolution (See Fig. \ref{carbon}, lower-right panel). The abundances inferred at full and reduced are within $<0.05$\,dex, and some CH features are now detectable by eye. \citet{pat18} found A(C)$=6.40\pm0.2$ consistent with our value, and \citet{fre18} derived A(C)$=5.89\pm0.2$ but with an effective temperature $\sim170$\,K lower. The three values are in a reasonable agreement and match with the first upper limit derived by \citet{agu18II}, A(C)$<6.3$, using the medium-resolution ($R\sim2,500$) ISIS spectrum.

\begin{table}
\begin{center}
\renewcommand{\tabcolsep}{5pt}
\centering
\caption{Absolute abundances for individual lines.
\label{lines}}
\begin{tabular}{cccccc}
\hline
Element &$\lambda$ (\AA)&$\log gf (\rm dex)$& EW(m\AA) &$\mbox{A(X)}$  \\
 \hline                                     
Li I   & 6707.8 & -- & 19.3$\pm$1.1 & 2.02 \\	    
C (CH) & G-band & -- & -- &6.20 \\	    
Na I   & 5889.95 & 0.11 &9.2$\pm$1.0&1.94 \\	  
Na I   & 5895.92 &-0.18 &12.9$\pm$0.9&2.22 \\	  
Mg I   & 3829.35 &-0.23 &79.7$\pm$2.0&4.66 \\   
Mg I   & 3832.30 & 0.12 &83.3$\pm$2.0&4.50 \\	  
Mg I   & 3838.29 & 0.39 &92.2$\pm$2.0&4.63 \\	  
Mg I   & 5167.32 &-1.03 &63.7$\pm$1.2&4.78 \\	  
Mg I   & 5172.68 &-0.40 &89.2$\pm$1.4&4.54 \\	  
Mg I   & 5183.60 &-0.18 &94.7$\pm$1.2&4.48 \\	  
Al I   & 3944.00 &-0.62 &24.2$\pm$1.4&2.36 \\	  
Al I   & 3961.52 &-0.32 &24.3$\pm$1.4&2.34 \\	  
Si I   & 3905.52 &-1.09 &36.7$\pm$1.8&4.05 \\   
Ca II  & 3933.66 & 0.13 &92.82$\pm$1.9&0.68 \\	  
Ca II  & 3968.47 &-0.16 &55.9$\pm$2.3&0.65 \\	  
Fe I   & 3859.92  &-0.68 & -- &$<$2.02 \\
   \hline
\end{tabular}
\end{center}
\end{table}

\subsection{Other elements}
J0023+0307 exhibits relatively high abundances of some elements compared to iron or calcium. The spectrum presents strong \ion{Mg}{1} absorptions in the blue ($\sim383$~{nm}) and also in the magnesium triplet \ion{Mg}{1}b at 516--518~{nm}. Six very clear features (See Table~\ref{lines}) have been identified and fitted giving an abundance of $A(\rm Mg)=4.60\pm0.11$ corresponding to [Mg/H]$=-2.93$. 
Si also shows an absorption at ($390.5$~{nm}) leading us to derive $A(\rm Si)=4.05\pm0.05$ ([Si/H]$=-3.46$). The shape of the aluminum features at $394.4$~{nm} and $396.1$~{nm} is not so well defined, however the derived abundances are the same, $A(\rm Al)=2.35\pm0.08$ ([Al/H]$=-4.02$). The Calcium $H\&K$ analysis is significantly improved compared to that presented in \citet{agu18II}. Both resonance lines are resolved and a complex ISM structure is present (See Fig. \ref{carbon}, upper-left panel). We find $A(\rm Ca)=0.66\pm0.09$ ([Ca/H]$=-5.65$), which is $\sim0.5$\,dex higher than \citet{agu18II}, from significantly lower resolution ($R\sim 2500$) spectrum; and fully compatible with \citet{pat18,fre18}. 
Assuming a [Ca/Fe]$\geqslant0.40$, we use the calcium abundance to set a metallicity upper-limit of [Fe/H]$\leqslant -6.1$. Finally the two sodium lines around $\sim589.2$~{nm} present broader line profiles than expected and moderate quality. The best FERRE fits allow us to derive $A(\rm Na)=2.08\pm0.21$ ([Na/H]$=-4.09$). Some derived upper limits are shown in table.

\begin{table*}
\begin{center}
\renewcommand{\tabcolsep}{2pt}
\centering
\caption{Abundances for individual species J0023+0307.\label{results2}}
\begin{tabular}{ccccccccccccccccccccc}
\hline\hline
 &  & \multicolumn{4}{c}{This work} & & \multicolumn{4}{c} {Fran\c{c}ois et al. (2018)}  & & \multicolumn{4}{c} {Frebel et al. (2018)}\\
\cline{3-6} \cline{8-11} \cline{13-16}
Species & $\log\epsilon_{\odot}$\,(X)$^1$ & $\log\epsilon$\,(X) & $\mbox{[X/H]}^1$ & $\sigma$ & $N$ &
                                      & $\log\epsilon$\,(X) & $\mbox{[X/H]}^2$ & $\sigma$ & $N$ & &$\log\epsilon$\,(X) & $\mbox{[X/H]}^2$ & $\sigma$  & $N$ \\
\hline
Li I    & 1.05 &     2.02 &             & 0.05    & 1 &&          &          &         &             &&   1.70   &         &0.20  &1 &\\
CH      & 8.39 &     6.20 &  $-$2.19    & 0.20    &   &&     6.40 &  $-$1.99 & 0.30    &             &&   5.89   & $-$2.50 &0.20  &  &\\
Na I    & 6.17 &     2.08 &  $-$4.09    & 0.21    & 2 &&          &          &         &             &&   2.24   & $-$3.93 &0.30  &2 &\\
Mg I    & 7.53 &     4.60 &  $-$2.93    & 0.11    & 7 &&     5.10 &  $-$2.43 & 0.23    & 3           &&   4.63   & $-$2.90 &0.10  &8 &\\
Al I    & 6.37 &     2.35 &  $-$4.02    & 0.12    & 2 &&          &          &         &             &&   2.30   & $-$4.07 &0.20  &2 &\\
Si I    & 7.51 &     4.05 &  $-$3.46    & 0.10    & 1 &&     4.20 &  $-$3.31 & 0.23    & 1           &&   3.94   & $-$3.57 &0.20  &1 &\\
Ca II   & 6.31 &     0.66 &  $-$5.65    & 0.09    & 2 &&     0.60 &  $-$5.71 & 0.20    & 1           &&   0.57   & $-$5.74 &0.20  &1 &\\         
Ti II   & 4.95 &  $-0.42$ &  $-$5.37    & 0.23    & 15&&          &          &         &             &&   $<$0.6 & $<-$4.35&      &  &\\
Cr I    & 5.64 &  $<$1.54 &  $<-$4.1    &         &   &&          &          &         &             &&   $<$1.4 & $<-$4.24&      &  &\\
Fe I    & 7.45 &  $<1.95$ &  $<-5.5$    &         &   &&    $<$3.5& $<-$3.95  &         &             &&   $<$1.9 & $<-$5.55 &      &  &\\
Ni I    & 6.23 &  $<$2.93 &  $<-$3.3    &         &   &&          &          &         &             &&   $<$2.2 &  $<-$4.03 &     &  &\\
Sr II   & 2.92 &  $<-$1.68&  $<-$4.6    &         &   &&    $<$0.0& $<-$2.92 &         &             &&   $<-$1.5&  $<-$4.42 &     &  &\\
Ba II   & 2.17 &  $<-$1.33&  $<-$3.5    &         &   &&          &          &         &             &&   $<-$1.3&  $<-$3.47 &     &  &\\
\hline
\end{tabular}
\end{center}
 $^1$Solar abundances from \citet{asp05}\\
 $^2$Solar abundances converted to  \citet{asp05}\\
 
\end{table*}

\subsection{Lithium}
\begin{figure}
\begin{center}
{\includegraphics[width=80 mm, angle=0]{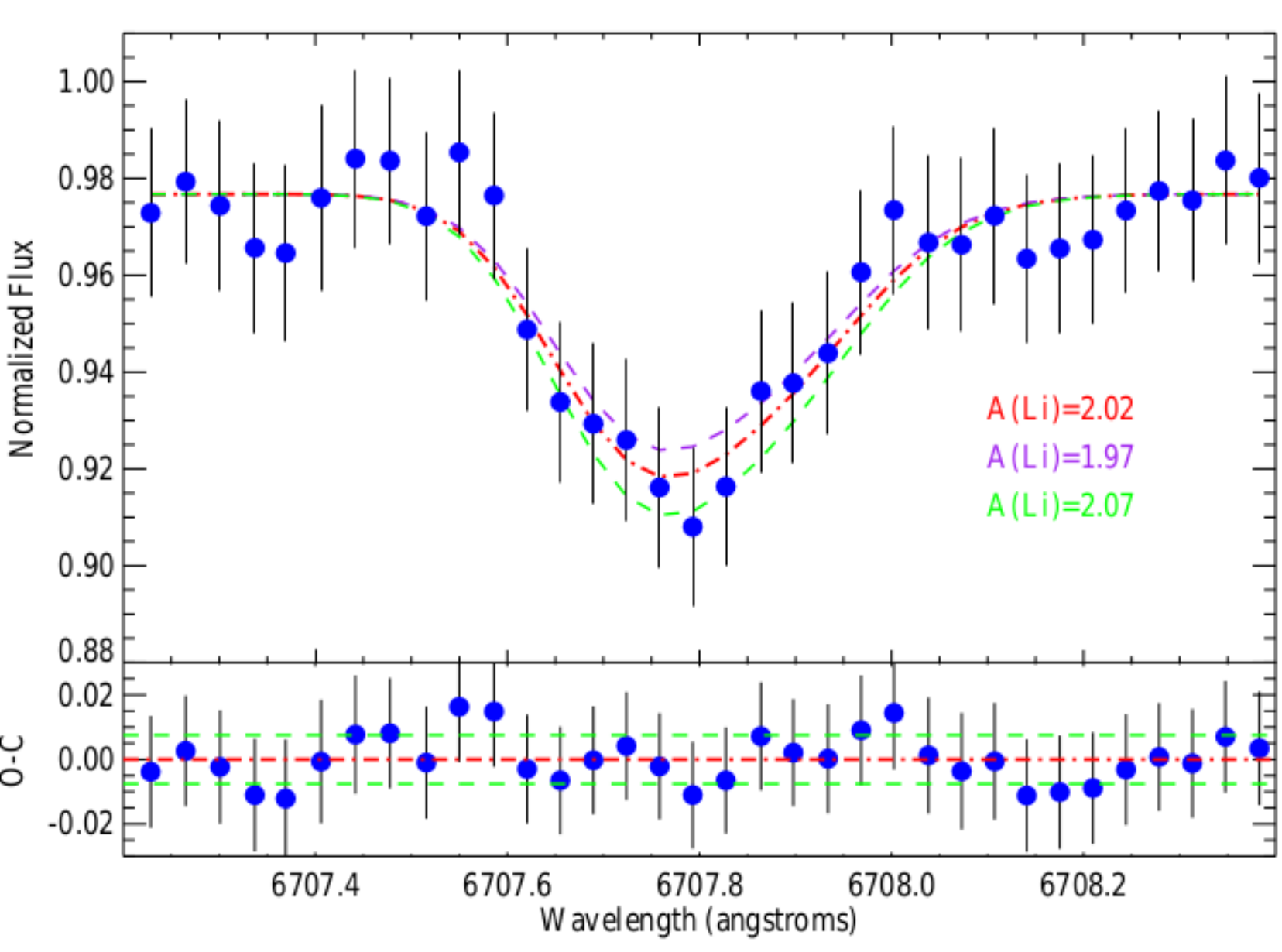}}
{\includegraphics[width=80 mm, angle=0]{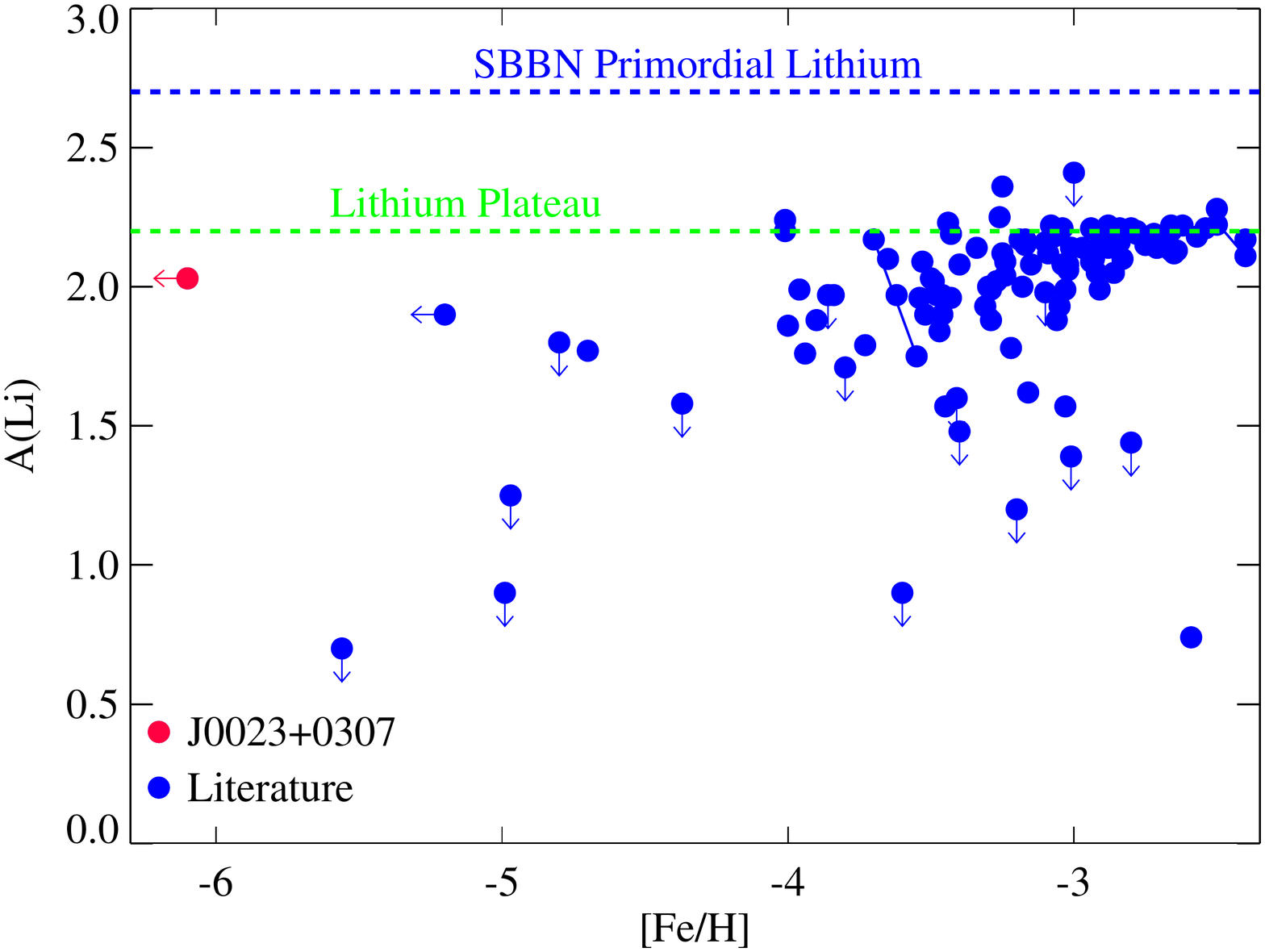}}
\end{center}
\caption{{\it Upper panel}: Li doublet of J0023$+$0307 rebinned to 0.035{\AA}/pixel (1.6~km~s$^{-1}$/pixel) together with the best fit model (A(Li)$=2.02$, $\sigma = 0.05$) and the fit residual, providing a S/N$\sim130$. We also show for comparison two additional synthetic spectra at about 3-$\sigma$ from the best fit.} 
{\it Lower panel}: Li abundance, A(Li), versus metallicity, $\left[{\rm Fe/H}\right]$, of J0023$+$0307 compared with other dwarf - turn-off stars ($\log g \geq 3.7$)  with Li abundance values from \citet{boni18} and references therein. %(\citet{asp06II,fre08,aoki09,mel10,sbo10,caff11,sbo12,mas12,boni12,han14,han15I,mat17,boni18}).
Blue filled circles connected with a solid line indicates the spectroscopic binaries in \citet{jon08b,aoki12}. The Lithium plateau (also called “Spite Plateau”) reference is shown as solid line at a level of A(Li)~$=2.20$\,dex. Blue dashed line represents the primordial lithium value (A(Li)$\sim$2.7) from WMAP \citep{spe03}.
\label{lithium}
\end{figure}

We have been able to clearly detect the Li doublet (see Fig.~\ref{lithium}) in the rebinned spectrum of the extremely iron-poor star J0023$+$0307. We measure a S/N~130 with a bin of 0.035{\AA}/pixel near the Li doublet.
We fit the lithium profile with our automated fitting tool (see Section~\ref{obs}), using a grid of model spectra for three different Li abundances (A(Li)$=$2.4, 2.0, 1.6~dex), and with A(Li), continuum location and global shift of the line as free parameters, and we fix the global FWHM to 7.4~km/s (see Section~\ref{obs}). We derive an EW(Li)$=19.3\pm1.1$~m{\AA} (see Table~\ref{lines}).
To evaluate the statistical uncertainty, we perform a MonteCarlo simulation with 10,000 realizations by injecting noise, corresponding to the actual S/N ratio near the Li line, in the best-fit synthetic spectrum. We measure a Li abundance in the star J0023$+$0307 of A(Li)$=2.02\pm0.08$, including the error due to the uncertainty of the adopted $T_{\rm eff}$ \citep[see e.g.][]{jon08b}. 
The difference of $\sim 0.3$~dex in Li abundance with respect to the value reported in \citet{fre18} is not only related to our hotter adopted $T_{\rm eff}$. A 160K lower $T_{\rm eff}$ translates into about 0.1 dex lower Li abundance, so there remains still 0.2~dex difference to be explain. We believe that our better quality data allow us to provide a robust Li abundance determination.

Our Li abundance measurement is almost at the level of the Lithium Plateau but at metallicity level of [Fe/H]$<-6$, which is significantly lower than any previous measurement. In Fig~\ref{lithium} we also display the lithium abundances of extremely metal-poor unevolved stars including main-sequence and turn-off field stars with available Li measurements or upper limits. \citet{mat17II} suggested that no stars with [Fe/H]$< -3.5$ show Li abundances at the level of the Lithium plateau (A(Li)$=2.2\pm0.1$), apart from three stars \citep{jon08b,boni18}.
%, but still at metallicities [Fe/H]$\geqslant-4$. 
Recently, \citet{boni18} have been able to measure the Li abundance of the iron-poor star J1035$+$0641 (A(Li)$=1.9$, [Fe/H]$<-5.2$). Both stars are suggested to be CEMP-no stars, but J0023$+$0307 has roughly 1~dex lower C abundance. Other unevolved CEMP stars at [Fe/H]$<-4.5$ have either no Li detected or A(Li)$< 1.8$, which may suggest that Li is not connected with the C content~\citet{mat17II}.

\section{Discussion and conclusions}\label{conclusion}
J0023$+$0307 is a dwarf star with [Fe/H]$<-6.1$, A(Li)$=2.02$ and very large overabundances of carbon ([C/Fe]$>3.9$), magnesium ([Mg/Fe]$>3.2$) and silicon ([Si/Fe]$>2.6$). These chemical abundance ratios provide a strong argument against ISM pollution by a Pair-Instability Supernova (PISN), since similar shortages of Ca-Fe and Mg-Si should be expected. In particular \citet{tak18} found the ratios [Na/Mg]$\sim1.5$ and [Ca/Mg]$ \sim0.5-1.3$ as the most robust discriminant of PISN. For J0023$+$0307, both [Na/Mg]$=-1.1$ and [Ca/Mg]$=-3.2$, are quite far from those values. The low Fe abundance of J0023$+$0307 is fully compatible with a Pop III supernova scenario. The high [C/Fe] together with Mg and Si derived abundances may point out to fall-back \citep{ume03}, returning Fe and Ca to the black hole. %\citet{fre18} also mentioned material from stellar wind coming off a rotating massive first star could happened.
J0023$+$0307 is in the low band of absolute carbon abundance \citep{boni15,yoo16} and does not show radial velocity variations, discarding the scenario of mass transfer from a binary companion.
\citet{til18I} present complementary criteria to disentangle different second generation star formation scenarios, mono-enriched or multi-enriched. The proposed methodology uses two different criteria, a semi-analytical model to determine the formation sites and other based on the divergence of the chemical displacement (DCD), for further details see \cite{til18I,til19}. In particular, the chemical signatures of J0023$+$0307 may fit well in the most likely mono-enriched area of the [Mg/C] vs [Fe/H] diagram presented in \citet{til18I}. SDSS J1035$+$0641 (with a metallicity of [Fe/H]$<-5.2$) discovered by \citet{boni15} also presents a high probability of being a second generation mono-enriched star \citep{til19}. 

 \citet{boni18} has recently detected lithium (A(Li)$=1.9$) in J1035$+$0641 close to the Lithium Plateau. J0023$+$0307 with Li abundance of A(Li)$=2.02$ surprisingly nearly recovers the same level of the Lithium Plateau at about 1 dex less iron content. The presence of lithium in this extremely iron-poor star at [Fe/H]$\lesssim -6$ reinforces the production of lithium at the Big Bang, and places a stringent constraint to any theory aiming at explaining the cosmological Li problem. The fact that no star in this large metallicity regime ($-6<$[Fe/H]$<-2.5$) has been detected to show a Li abundance between that inferred from SBBN and the Li plateau, makes this upper boundary of Li abundance (or extended Li plateau) at low metallicities difficult to explain by destruction in the stars, and may support a lower primordial Li production, driven by non-standard nucleosynthesis processes.

\begin{acknowledgements}
DA thanks the Leverhulme Trust for financial support.
DA acknowledges the Spanish Ministry of Economy and Competitiveness (MINECO) for the financial support received in the form of a Severo-Ochoa PhD fellowship, within the Severo-Ochoa International PhD Program.
DA, JIGH, CAP, and RR also acknowledge the Spanish ministry project MINECO AYA2017-86389-P. JIGH acknowledges financial support from the Spanish Ministry of Economy and Competitiveness (MINECO) under the 2013 Ram\'on y Cajal program MINECO RYC-2013-14875. 
\end{acknowledgements}

\end{document}